# PERFORMANCE IMPROVEMENT OF LANGUAGE-QUERIED AUDIO SOURCE SEPARATION BASED ON CAPTION AUGMENTATION FROM LARGE LANGUAGE MODELS FOR DCASE CHALLENGE 2024 TASK 9

Technical Report

*Do Hyun Lee[1], Yoonah Song[1], and Hong Kook Kim[1,2,3]*

[1] AI Graduate School, [2] School of EECS
Gwangju Institute of Science and Technology, Gwangju 61005, Republic of Korea
[3] Aunion AI, Co. Ltd., Gwangju 61005, Republic of Korea
{zerolee12@gm., yyaass0531@gm., hongkook@}gist.ac.kr

## ABSTRACT

We present a prompt-engineering-based text-augmentation approach applied to a language-queried audio source separation (LASS) task. To enhance the performance of LASS, the proposed approach utilizes large language models (LLMs) to generate multiple captions corresponding to each sentence of the training dataset. To this end, we first perform experiments to identify the most effective prompts for caption augmentation with a smaller number of captions. A LASS model trained with these augmented captions demonstrates improved performance on the DCASE 2024 Task 9 validation set compared to that trained without augmentation. This study highlights the effectiveness of LLM-based caption augmentation in advancing language-queried audio source separation.

## 1. INTRODUCTION

Language-queried audio source separation (LASS) is the task of extracting sound sources using textual descriptions, also referred to as 'separate what you describe' [1], [2]. This approach allows users to separate specific audio sources through natural language instructions. The ability to separate desired sources using only natural language descriptions enables potential applications such as simplifying audio editing [3] and improving multimedia content retrieval [4]. However, two major challenges related to data availability and quality are encountered in training deep learning models for this task.

First, multiple descriptions can exist for one audio clip. For example, an audio clip of a dog barking could be described as "a dog is barking loudly" or "a loud bark from a dog is heard." Second, annotating audio with captions is challenging and costly, which limits dataset size. To address data scarcity, the most intuitive and effective solution is to augment caption data. This approach allows us to obtain various descriptions for a single audio clip while also mitigating the issue of limited annotated data.

Previous studies such as [5] and [6] utilized AudioSet [7], which is used for sound event classification and only provides sound class types without detailed descriptions. However, label information is insufficient for describing relationships between multiple sound events, such as their spatial and temporal relationships, necessitating the use of sentence-like descriptions. Consequently, the study in [2] used ChatGPT to generate and filter captions and then incorporate them for training with AudioSet for LASS. In practice, datasets like Clotho [8] and WavCaps [9] that are commonly used for audio captioning are utilized for training LASS models. These datasets were enhanced by using ChatGPT, a large language model (LLM), to generate useful captions.

In our work, we utilize large language models (LLMs) to augment captions, aiming to improve the performance of LASS models without increasing the number of annotated audio samples. We examine the effect of various input prompts on LASS performance to discover the prompt that yields the best performance by augmenting the training dataset. Consequently, this augmented dataset was then added to train the baseline model for DCASE 2024 Task 9 [10]. As a result, we achieved an SDR of 7.69 dB on the validation set, improving upon the baseline's SDR of 5.70 dB.

The remainder of this report is organized as follows. Section 2 provides a detailed explanation of our caption augmentation method using LLMs. Section 3 describes the datasets used for training the LASS model and presents our experimental results and discussion. Finally, Section 4 concludes the report.

## 2. METHODS

### 2.1. Prompt engineering

To enhance the LASS performance, a prompt engineering approach is introduced here to design and refine prompts for generating texts from language models. By carefully crafting prompts to elicit desired responses from the model, it is expected to produce more accurate and contextually appropriate captions. Notice that the captions in WavCaps [9] and Clotho [8], which were used for the baseline of DCASE Challenge Task 9, were generated by ChatGPT with some proper prompts.

---
[*] D. H. Lee and Y. Song equally contributed.
[**] This work was supported in part by the GIST-MIT Research Collaboration grant and by the "Practical Research and Development support program supervised by the GTI(GIST Technology Institute)" grant funded by the GIST in 2024. In addition, it was supported by "Project for Science and Technology Opens the Future of the Region" program through the Innopolis Foundation funded by Ministry of Science and ICT (Project Number: 2022-DD-UP-0312). In addition, it was supported by Artificial intelligence industrial convergence cluster development project funded by the Ministry of Science and ICT(MSIT, Korea)&Gwangju Metropolitan City.

Table 1: List of prompts and exemplar responses of three different prompt types using the large language model phi-2.0.

| Prompt Type | Instruction (Prompt) | Response |
|---|---|---|
| Simple Prompt | Generate unique (#) captions for the following sounds, ensuring each description varies distinctly from the others: {Caption} | ① The cacophony of a TV channel gone awry, a symphony of static and distortion. ② A disjointed chorus of electronic chaos, a dissonant melody of broken signals. ③ A jumbled orchestra of audio mishaps, a cacophony of static and interference. ④ A disarrayed symphony of electronic disarray, a discordant chorus of broken channels. |
| Modification of Clotho Prompt | Generate (#) captions for the following sounds. Use a subject-verb-object grammatical structure, do not use the word 'heard,' do not describe the temporal order of the sounds, and ensure that each caption is less than 20 words. | ① The TV's audio is distorted. ② The TV's sound is garbled. ③ The TV's sound is unclear. ④ The TV's sound is muffled. |
| Modification of WavCaps Prompt | Generate (#) captions for each sound. Use a subject-verb-object structure. Remove all references to specific times, locations, devices, and named entities—replace people names with 'someone.' Summarize sound events in no more than 20 words per caption. Avoid using 'heard' or recording' specifics. Start each caption with its index and output 'Failure' if the description is not directly about the sound. | ① The TV is making a strange noise. ② The TV is not working properly. ③ The TV is malfunctioning. ④ The TV is not functioning. |

WavCaps and Clotho used a prompt each to generate a sentence by only considering the event label. It is evident that these prompts might not be suitable for our task because we need to augment sound description captions in a sentence level not event word level. To remedy this issue, we need to redesign a prompt that can consider a sentence with a similar meaning to original prompts.

Table 1 shows three different prompts that are investigated in this work. As shown in the table, we first designed a "Simple Prompt" to generate similar captions that are given in the baseline datasets. Next, we modified the prompt for the Clotho dataset so the prompt could give an emphasis on a sentence level concentration, which is referred to as "Modification of Clotho Prompt." Similarly, we also modified the prompt for WavCaps, which is referred to as "Modification of WavCaps Prompt." Note here that to generate captions with proposed prompts, the phi-2.0 LLM [11] was used, which was a 2.7 billion-parameter language model released by Microsoft and had better language understanding and generation performance than Llama-7B [11]. Despite its smaller size, it was found that the phi-2.0 LLM was adequate for our task.

The third column of Table 1 shows the captions generated with each prompt shown in the second column of the table. For "Simple Prompt," the absence of specific sentence structure or length guidelines resulted in often lengthy and contextually inconsistent sentences. On the other hand, the generated captions with the Modified Clotho Prompt became shorter and simpler in structure and more consistent than those with Simple Prompt. However, this approach also led to several duplicated captions. Next, the Modified WavCaps Prompt generated as many sentences as the LLM could generate. In other words, even if five sentences were requested to generate, the LLM with Modified WavCaps Prompt resulted in "Failure" if the description was not directly related with the sound.

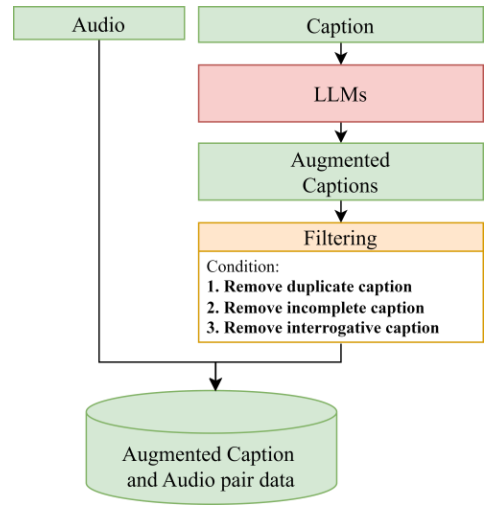

Figure 1: Flowchart of the caption augmentation process using LLMs.

Table 2: Performance comparison of different types of prompts applied to the Clotho v2 Dataset.

| Prompt Type | SDR | SDRi | SI-SDR |
|---|---|---|---|
| Baseline (no augmentation) | 3.079 | 3.044 | 1.105 |
| Simple Prompt | 3.011 | 2.976 | 1.295 |
| Modification of Clotho Prompt | 3.133 | 3.098 | 1.361 |
| Modification of WavCaps Prompt | 3.320 | 3.285 | 1.505 |

All three types of prompts were designed to generate four captions. However, we observed instances of duplicated or incomplete captions in our results. To address these issues, as shown in Figure 1, we implemented a filtering process. Consequently, not all wav files were augmented with four captions as initially intended.

### 2.2. Prompt selection via performance comparison

To select an appropriate prompt for DCASE 2024 Challenge Task 9, we carried out performance evaluation between the three prompts described in Section 2.1. Table 2 compares the SDR, Sri and SI-SDR according to different types of prompts. The Baseline is the LASS model by using randomly selected one caption out of five captions given to each audio file. Next, we the LASS model using the captions with Simple Prompt. As shown in the table, Simple Prompt lowered SDR and SDRi, compared to the Baseline. This was because the generated captions by Simple Prompt were found to be lengthy and contextually inconsistent. This inconsistency introduced noise, which negatively impacted performance.

In contrast, Modification of Clotho Prompt and Modification of WavCaps Prompt outperformed Simple Prompt, since the captions generated by them likely included less contextual noise than those by Simple Prompt. In particular, Modification of WavCaps Prompt achieved the highest performance in SDR, SDRi and SI-SDR. Based on this performance comparison, we chose to augment our data using Modification of WavCaps Prompt.

## 3. EXPERIMENTS

### 3.1. Dataset

The development set provided for the DCASE 2024 Challenge Task 9 baseline system consisted of audio samples from the Clotho v2 and FSD50K datasets. The captions for each audio sample in the datasets were originally created and released by using ChatGPT to refine raw descriptions.

In addition to the two datasets, WavCaps was also employed in this work. Due to the Challenge Rules for Task 9, we excluded audio samples from FreeSound. For details on the datasets we used, refer to Table 3.

1) *FSD 50k*: Freesound Dataset 50k (FSD50k) [10] was a comprehensive collection of human-labeled sound events, comprising 51,197 Freesound clips. Each audio clip was classified into one of 200 AudioSet Ontology classes. These included human noises, sounds of objects, animal sounds, natural sounds, musical instruments, etc.

Table 3: Summary of training datasets.

| Category | Dataset | Num. clips | Num. captions |
|---|---|---|---|
| Baseline Dev Set | FSD 50k | 40,966 | 40,966 |
|  | Clotho | 3,839 | 19,195 |
| WavCaps | BBC sound | 31,201 | 31,201 |
|  | Soundbible | 1,232 | 1,232 |
|  | AudioSet | 108,317 | 108,317 |

2) *Clotho v2*: Clotho v2 [8] was an audio captioning dataset comprising 5,929 audio clips. The dataset was divided into 3,839 audio clips for development, 1,045 for validation, and 1,045 for evaluation. Each clip was accompanied by five human-annotated captions, each containing between 8 to 20 words.

3) *WavCaps*: WavCaps [9] was a large-scale dataset of weakly-labeled audio captions. It consisted of 403,050 paired-caption audio clips, whose length was approximately 7,568 hours in total. The audio clips were excerpted from FreeSound, BBC Sound Effects, SoundBible, and AudioSet, where their captions were generated by ChatGPT from raw audio descriptions.

### 3.2. Model training

We adopted the baseline system of DCASE 2024 Task 9 for model training. The baseline system consisted of two main components: the query net, which extracts embeddings for language queries, and the separation net, which performs source separation. For the query net, we utilized contrastive language-audio pre-training (CLAP). For the separation net, we used ResUNet, an advanced version of Unet widely employed in source separation tasks. We used the Adam optimizer with a learning rate of $1 \times 10^{-3}$ and a batch size of 64, and trained the model for 100 epochs. Additionally, we incorporated a pre-trained checkpoint provided by AudioSep, which was trained on a larger dataset and demonstrated better performance. Finally, we fine-tuned the pre-trained model using our caption-augmented datasets to enhance performance.

### 3.3. Discussion

Performance was evaluated using the metrics defined in DCASE 2024 Challenge Task 9 [4], i.e., signal-to-distortion ratio (SDR), signal-to-distortion ratio improvement (SDRi), and scale-invariant signal-to-distortion ratio (SI-SDR). Table 4 compares the performance of different LASS models trained using various combinations of training datasets with/without caption augmentation on the validation dataset of DCASE 2024 Challenge Task 9.

As shown in the table, the 'Baseline' model trained on the baseline dev set achieved an SDR of 5.817 dB, which was similar to the checkpoint SDR of 5.708 dB provided by the baseline system. We then augmented the baseline dev set with captions generated by Modification of WavCaps Prompt, resulting in a significant increase in SDR to 6.547 dB. Such substantial improvement was due to significantly increased the total number of captions. Next, when we incorporated the WavCaps dataset with generated captions, excluding FreeSound, we observed the further SDR increase from 7.500 to 7.750 dB.

Next, to further improve SDR performance, we utilized the pre-trained AudioSep model [2], which was trained on over 2 mil-lion clips from weakly labelled datasets such as AudioSet, VGG-Sound, and AudioCaps. The pre-trained AudioSep model achieved higher SDR than the Baseline LASS model that was even trained with augmented captions. This result inspired us to finetune the AudioSep model with augmented captions. Consequently, the fine-tuned AudioSep model provide the highest SDR among all the single models, as shown in the table.

Table 4: Performance comparison of different LASS models trained using various combinations of training datasets with/without caption augmentation on the validation dataset of DCASE 2024 Challenge Task 9.

| No. | Model | Training dataset | Training Approach | Caption Augmentation | SDR | SDRi | SI-SDR |
|---|---|---|---|---|---|---|---|
| 1 | Baseline | Baseline Dev Set | Full | ✗ | 5.817 | 5.782 | 3.837 |
| 2 | Baseline | Baseline Dev Set | Full | ✓ | 6.547 | 6.512 | 4.636 |
| 3 | Baseline | Baseline Dev Set + WavCaps | Full | ✗ | 7.500 | 7.465 | 5.795 |
| 4 | Baseline | Baseline Dev Set + WavCaps | Full | ✓ | 7.750 | 7.715 | 6.161 |
| 5 | AudioSep | - | Pretrained | - | 8.195 | 8.160 | 6.708 |
| 6 | AudioSep | Baseline Dev Set + WavCaps | Fine-tuning | ✗ | 8.370 | 8.335 | 7.109 |
| 7 | AudioSep | Baseline Dev Set + WavCaps | Fine-tuning | ✓ | 8.459 | 8.424 | 7.072 |
| 8 | Ensemble (4+5+6+7) | - | - | - | 8.599 | 8.564 | 7.497 |
| 9 | Ensemble (3+4+5+6+7) | - | - | - | 8.610 | 8.575 | 7.493 |

Finally, we ensembled the models with a weighted sum method. As shown in the eighth row of Table 4, the first ensemble model combined with the models 4, 5, 6 and 7 resulted in the highest SI-SDR of 7.497, which was a significant improvement compared to the baseline. Next, we constructed the second ensemble model combined with the models 3, 4, 5, 6, and 7. Consequently, we achieved the highest SDR among all the single and ensemble models in the table, but a slightly reduced SI-SDR, compared to the ensemble model.

## 4. CONCLUSION

This report presented submitted LASS models applied to DCASE 2024 Challenge Task 9. Particularly, we focused on caption augmentation via a prompt engineering approach. In other words, we modified prompts by taking into account audio scene sentences and generated five or less sentences by giving a selected prompt to LLM. We evaluated the SDR performance of the baseline and AudioSep models with/without caption augmentation. Consequently, it was shown that the caption augmentation by the proposed prompt increased SDR without regard to the backborn of LASS model, such as the baseline and AudioSep. Moreover, the ensemble model was finally shown to be the best performance on the DCASE 2024 Task 9 validation set.